# Symmetry-aware SFC Framework for 5G Networks

Hajar Hantouti*, Nabil Benamar*, Miloud Bagaa‡ and Tarik Taleb‡

*Moulay Ismail University of Meknes, Morocco

‡Aalto University, Espoo, Finland

*Abstract*—Network Function Virtualization (NFV), network slicing, and Software-Defined Networking (SDN) are the key enablers of the fifth generation of mobile networks (5G). Service Function Chaining (SFC) plays a critical role in delivering sophisticated service per slice and enables traffic traversal through a set of ordered Service Functions (SFs). In fully symmetric SFCs, the uplink and downlink traffic traverse the same SFs, while in asymmetric SFC, the reverse-path may not necessarily cross the same SFs in the reverse order. Proposed approaches in the literature support either full symmetry or no symmetry. In this paper, we discuss the partial symmetry concept, that enforces the reverse path to traverse the SFs only when needed. Our contribution is threefold. First, we propose a novel SFC framework with an abstraction layer that can dynamically create partial or full symmetric SFCs across multiple administrative and technological cloud/edge domains. According to the Key Performance Indicators (KPIs) and desired objectives specified at the network slice intent request, the abstraction layer would automatise different SFC operations, but specifically generating partial or full symmetric SFCs. Second, we propose an algorithm to dynamically calculate the reverse path for an SFC by including only SFs requiring symmetry. Third, we implement a prototype application to test the performance of the partial symmetry algorithm. The obtained results show the advantages of partial symmetry in reducing both the SFC delivery time and the load on VNFs.

## I. INTRODUCTION

Service Function Chaining (SFC) has recently gained significant interest from academia and industry and plays a crucial role in automating network services deployment. The SFC concept leverages traffic engineering techniques in forwarding traffics between peers. Owing to new networking concepts like SDN, NFV, and network slicing new SFC techniques have been suggested for the next-generation networks to support the management of complex services. [1]. Moreover, the deployment of dynamic SFC within the next generation networks has empowered SFC use cases and enhanced its flexibility.

SFC is defined as a set of operations enabling traffic steering through a set of ordered SFs, which can be a form of Virtual Network Functions (VNF) or another form of Physical Network Function (PNF). An SF can be a Firewall, an Intrusion Prevention System (IPS), or a Network Address Translation (NAT) to cite a few. An example of NFs in the 5G network includes Access and Mobility Management Function (AMF), Authentication Server Function (AUSF), Security Edge Protection Proxy (SEPP), and Service Communication Proxy (SCP) [2].

With the growing number of users and devices, network providers have been forced to reform their infrastructure (i.e. using the NFV-SDN-Slicing technologies) to meet the required complex services with a constrained QoS. While 5G promises low latency and more flexibility, various critical/sensitive vectors are concerned. Healthcare and autonomous driving are examples of use cases that require customized services based on very low latency, high accuracy, and reliability. Other use cases like the Internet Of Things (IoT) or factories might prioritize cost reduction. Hence, as an infrastructureindependent technology, SFC is critical for managing and optimizing every 5G complex service customized based on various network features and subscriber preferences. Such a service customization level makes SFC solutions challenging compared to SFC deployed in 4G networks (fewer requirements). Moreover, the promise of latency reduction in 5G networks leads the researchers and industry to think about the latency of both upload and download traffic, while only the download traffic latency is considered in 4G.

As path asymmetry is the default rule on the Internet, some issues arise when nodes have to be crossed in both directions. For instance, a stateful firewall that does not see an outgoing SYN will not accept incoming SYN-ACK and will discard those packets. Likewise, if a NAT is on the path, the SF instance that handled the outgoing packet must be involved in the reverse path; otherwise, the connection will not be established. Other kinds of SFs need to ensure the consistency of flow state that should also be involved in both communication directions, such as proxies, IDS/IPS, NAT···etc.

For an SFC, traffic symmetry depends on the SFs requirements involved in the reverse path. The reversed direction defines the backward traffic from the target to the source, while the forward direction represents the traffic from the source to the destination. Different use cases permit to set up a reverse path by defining a different chaining order or using a bidirectional chain that defines fully symmetric traffic. However, these scenarios are not realistic, because some SFs may require the returning traffic to pass through them (i.e., requiring symmetry) while others may not. As a result, the network performance can be seriously reduced when forcing the traffic to pass by all the SFs in the reverse path (i.e., full symmetry). Moreover, an unnecessary load is added to the SFs, reducing their overall performance. To address this issue, the reverse-path should only visit the SFs when required (i.e., the SFs that require symmetry [3]).

In this paper, we propose an SFC framework for 5G system, with an abstraction layer that permits to dynamically create partial/full symmetric SFCs across multiple administrative and technological cloud/edge domains. According to the desired

Key Performance Indicators (KPIs) and the network slice intent request, the abstraction layer would generate either partial or full symmetric SFCs. This strategy offers the capability, cohability, and elasticity to network slice to meet the network requirements while reducing costs. Then, we propose an algorithm to calculate the correct reverse path for a given chain. A prototype implementation is realized to assess the performances of partial/full symmetry chains. The prototype is of a form of an SFC application that implements the proposed algorithm to calculate the reverse path, based on individual symmetry requirements of SFs. It dynamically inserts flow rules to Service Function Forwarders (SFFs) to steer traffic in the opposite direction. The application helps to enforce the reverse path. As a result, the delivery time is reduced (RTT, throughput and transfer). The rest of the paper is organized as follows. The concept of symmetry for SFC, context, and use cases are covered in section II. Section III describes the proposed SFC framework and the reverse path calculation algorithm. section IV presents an implementation prototype and evaluates the performance of our prototype, comparing partial and full symmetry SFCs. While Section V presents related work for traffic symmetry in SFC. Finally, section VI concludes the paper.

## II. Background on SFC and related technologies

### A. IETF SFC architecture

The SFC architecture, as described in RFC 7665, represents the creation, maintenance, and deployment of end-to-end SFCs in a network. It is based on topological independence from the underlying network topology [3] and deployment context. The architecture specifies logical components to compose an SFC overlay using encapsulation and specific traffic steering techniques.

*1) SFC logical components:* The architecture is composed of logical components that are responsible for different types of SFC operations including:

- <u>SFFs:</u> capable of forwarding traffic to and from the connected SFs based on the SFC information carried with the SFC encapsulation in the packets. The SFF maintains Service Function Path (SFP) (i.e., the set of SF-SFF associations creates an SFP) forwarding information needed for the traffic steering. It can also maintain the state in specific scenarios to ensure symmetry for example.
- <u>SFs:</u> can be any OSI layer function that permits to achieve specific treatment of packets. It can be seen as a resource for consumption as part of a composite service. An SF is connected to one or more SFFs from which it can receive or send data. For an SF to be part of the SFC architecture, it could be SFC-aware or connected to a proxy to adapt packets towards and from the SF to other SFFs.
- <u>CLs:</u> responsible for matching traffic flows against policy for specific SFC. As a result of the classification, the accurate SFC encapsulation is inserted into the packets and the relevant SFP is selected. The initial classification is achieved at the ingress of an SFC domain.
- Proxy: is a communication intermediate between the SF and SFF. It is responsible for attaching an SFCunaware SF to the SFF, it inserts the SFC encapsulation information to the traffic coming from the SF and strips the encapsulation information from the traffic going to the SF.

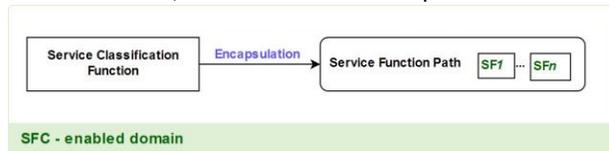

Fig. 1: High-level SFC Architecture inspired by RFC 7665.

*2) SFC encapsulation:* The encapsulation mechanism in SFC plays a crucial role in dynamic SFC deployment. It allows creating an SFC overlay and connecting the SFC components. Encapsulation is usually used in SFC in two ways: To enforce the SFP or to share the information between SFC components (Fig.1). The first type is referred to as a network encapsulation, where tunnels are used between the SFC components, different transport protocols can be used for this purpose (e.g. IPinIP, VxLAN, VxLAN-GPE or GRE). The second type is referred to as SFC encapsulation, where the SFC forwarding information is encapsulated in the packets and shared between the SFC components. Protocols like Network Service Header (NSH) [4] are used for this purpose, though the use of encapsulation in SFC solutions is not mandatory.

*3) SFC traffic steering:* Traffic Steering (TS) refers to the operations involved in forwarding traffic along the SFP. It can be achieved by newly defined protocols, such as NSH to share the SFC forwarding instructions between the SFC components. There are different TS techniques for SFC [1], which can be achieved using new SFC headers like NSH that carry SFC information. Such information reflects the instructions that should be applied by SFC components along the SFP. Other TS techniques involve existing packet headers or fields to encode the SFC information (e.g. MAC address, IP options, Vlan Id). Also, encapsulation methods can be used for SFC (e.g. MPLS, VxLAN).

### B. The link between SFC and next generation networks

Currently, SFC has become part of mobile networks, data centres, and broadband networks. When SFC is deployed in the SDN/NFV/Network slicing context, it allows composing customized services supporting fine granular policies in a dynamic and agile way. SFC allows avoiding strong adherence to the infrastructure and provides great deployment flexibility. Moreover, it ensures a dynamic service inventory, where SFs can be added or removed without breaking the chain. In NFV, virtual appliances of SFs could be deployed as VNFs and managed by the NFV operational components (e.g., NFV orchestrator, virtual infrastructure manager). Also, network slicing and SFC are becoming two technologies that go together. Network slicing allows the creation of independent, agile, and secure SFCs, while SFC enables the composition of added value services per slice. SDN plays a vital role in dynamically and automatically programming the forwarding
operations and controlling networking devices, such as the virtual switches that can be in the form of SFFs or CLs.

policy should be enforced. Otherwise, a partial symmetry strategy is enough and even recommended.

*2) Examples of SFs not requiring symmetry:* If an SFC contains SFs that require traffic communication in both directions and others not, then the partial symmetry policy can be applied. For example, a function that checks email for spam; it only needs to see the incoming content, not the control exchanges that drive it. Another example is a URL filter, which only needs to monitor the requests and not the response data. Also, a classification function does not require reverse traffic to pass through it, it needs to see traffic only in one direction but not both directions.

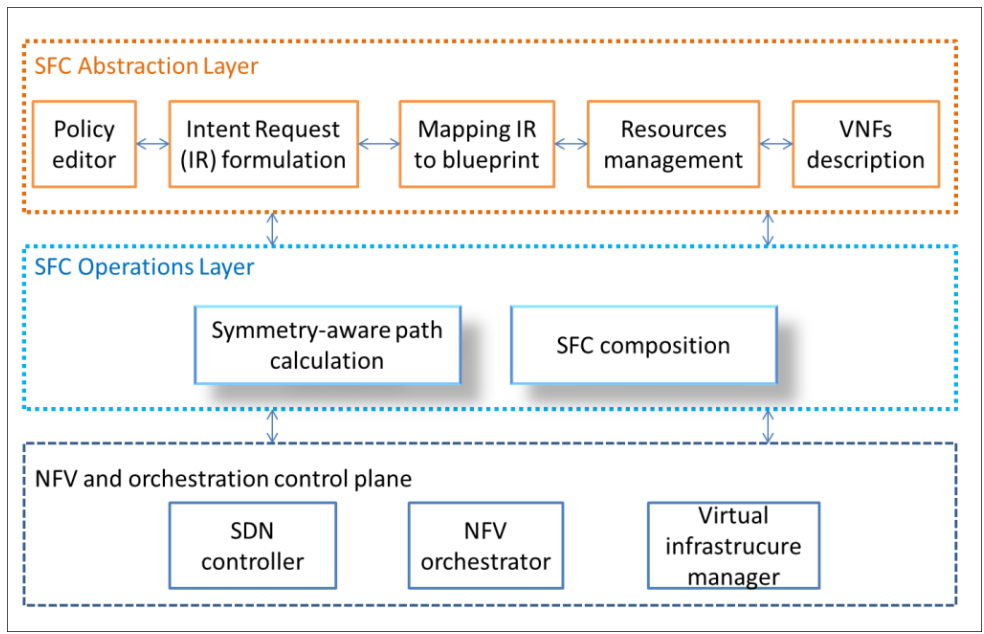

Fig. 2: Symmetry-aware framework for SFC in 5G networks.

*C. SFC symmetry scenarios*

SFCs can be either symmetric, asymmetric, or partially symmetric for different services and applications. Symmetric chains require the traffic to pass precisely by the reverse order of SFs in the opposite direction, while the asymmetric SFCs can pass by different SFs in the reverse path. In contrast to the previous chain types, the traffic in partial symmetry should pass through a small part of SFs in the reverse path.

In what follows, we give more detailed examples where the traffic communication symmetry may be required.

*1) Examples of SFs requiring symmetry:* The simplest case where symmetry is an issue is where one of the SFs is a stateful firewall. It is mandatory that the traffic for any specific flow passes through the same firewall instance in both forward and reverse directions. As an even more specific example, the firewall may only allow inbound TCP traffic for a given 4-tuple after seeing an outgoing TCP SYN for that 4-tuple. Another example is a TCP proxy, for similar reasons. Similarly, an IDS or a conventional firewall requires both communication directions before making adequate decisions. In case that all the SFs in an SFP should be stateful, then the full symmetry

*D. How can partial symmetry be deployed in 5G networks*

SFC partial symmetry is necessary for 5G networks to enhance the KPIs and offer enhanced SLAs for different reasons:

- Latency: The latency would be reduced due to the avoidance of SFs not requiring symmetry in the reverse path.
- Policy flexibility: The flexibility for managing path on both directions. In contrast to the traditional approaches, our proposed approach gives more freedom for managing the path on both directions by defining two different policies for the two directions. Having control over the reverse-path SFs traversal allows for more personalized SFCs and fine granular policies.
- Cost: Nowadays, a flexible consumption business model has been adopted widely in the cloud, where the telco provider pays as much as they use services (i.e., data traffic). For this reason, the adaptation of partial symmetry could reduce dramatically the cost. In fact, the VNFs do not have to treat unnecessary traffic.

## III. SYMMETRY-AWARE SFC FRAMEWORK

In this section, we first give insight into our framework proposition for 5G core network. Second, given the importance of partial symmetry SFC, we present an algorithm to calculate the reverse path of a chain based on symmetry requirements. The Algorithm is included in the symmetry-aware path calculation module of the proposed framework.

### A. Symmetry-aware SFC framework

The proposed framework supports the dynamic generation of SFCs across multiple administrative and technological cloud and edge domains. The framework is intended to be deployed in 5G core network and its primary goal is to automate the administration and management of SFC operations. Such operations can include but are not limited to, SFC composition and SFC path calculation. As such, the network administrator only supervises the framework without interacting in the SFC operations. First, the framework takes the customer request for instantiating an SFC automatically. Based on the network slice intent, blueprints of SFCs (also referred to as forwarding graphs) are generated based on the symmetry considerations of SFs. The SFs description will be loaded to the framework to specify the different SF requirements (i.g. symmetry). This information helps the SFC operation layer for taking the right decisions at various steps of SFC operations. This strategy offers the capability, co-habitability, and elasticity to meet the network requirements while reducing costs.

The framework runs on top of an NFV-SDN-sliced network that supports the framework operations. Each SFC can be deployed in a defined network slice, supporting fine-grained SFCs according to customer requirements. Being deployed in an NFV environment, the framework benefits from the NFV advancement to simplify the SF placement and SFC composition operations. The framework also benefits from SDN being deployed in a 5G core network via the SDN control plane to control the SFFs (basically, switches). Fig.2 shows a top-down overview approach of the proposed framework that consists of two consecutive layers. The framework orchestrate and manage an NFV and SDN based network through SDN controllers (SDN-C), NFV orchestrators (NFVO), and Virtual Infrastructure Manager (VIM), respectively.

The top layer, dubbed SFC abstraction layer, is responsible for the SFC and slices definition in an abstract manner. This layer has been defined as orthogonal on any underlying infrastructure. This layer is responsible for determining the slice blueprint and the communication with the bottom layer. Also, it consists of five modules, which are: policy editor, Intent Request (IR) formulation, mapping of IRs to blueprints, resource management and SFs definition module. Meanwhile, the bottom layer, dubbed SFC operations layer, implement all necessary functionalities for ensuring the life-cycle management of different SFCs starting from the instantiation, update, and deletion. After receiving the slice blueprint from the upper layer, this layer generates an instance of the blueprint by specifying various involved SFs, communication links, and required configurations. Accordingly, this layer communicates with different NFVOs, VIMs, and SDN-Cs to enforce the taken policies. New SFs could be created if needed. While Fig. 3 shows the detailed workflow for SFC intent deployment in our framework for a 5G network. First, the policy editor module prepares an Intent Request(IR) to be sent to the intent request formulation module, which interacts with other modules to prepare the intent deployment command. The IR formulation module starts by interacting with a module named "mapping IR to blueprints" to prepare the slice template for the intent which in our context represents an SFC. Afterwards, the IR formulation module requests the resources management module to calculate the required resources for the chain (resources are calculated according to the provided SLA). Later, the IR formulation module retrieves the SFs attributes (e.g. symmetry requirement). Consequently, the IR formulation module generates an intent deployment command that will be sent to the SFC operations layer for deployment by lower-level SFC applications.

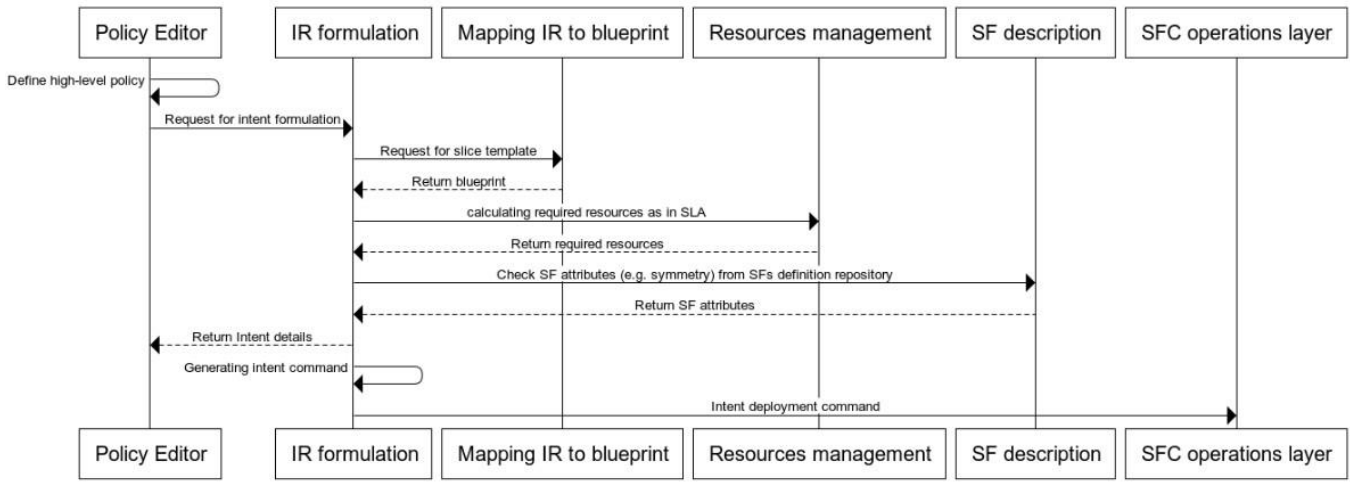

Fig. 3: Workflow for SFC intent deployment in the symmetry-aware SFC framework for 5G networks.

*1) SFC abstraction layer:* In what follows, we will describe further the components of the SFC abstraction layer.

   *a) Policy editor tool:* This module describes the intent independently of the SFC details. It mainly describes the service requirements and SLA.

   *b) Intent request formulation module:* We refer to intent as an object describing the added-value service. It can be defined by a policy-based application, a customer or an agent. It is presented in a readable human format to describe the requested service properties, such as SLA criteria. As an example of intent:
Intent Label: added-value-service1
Intent validity: 30 days
SLA: Bandwidth:x;Latency: y; Cost: z;

   *c) Mapping IR to blueprints:* A blueprint for network slice defines the structure of slice (i.e., the VNFs composing a slice), it can also refer to a service graph or SFC. This module is capable of translating the IRs into blueprints. Afterwards, the blueprints are used as templates to deploy the slices.

   *d) Resources management:* This module is responsible for managing resources (e.g. compute, storage, networking metrics). Once the SLA for the intent is defined, the resource management module calculates the accurate resources needed for a slice.

   *e) Definition of SFs:* This module allows the detailed definition of various attributes concerning a specific VNF or a group of VNFs. Such characteristics help as input for different operations' algorithms, such as the symmetry-aware path calculation or the SFC composition among others.

*2) SFC operations layer:* In this subsection, we describe the SFC operations layer, which contains different SFC applications responsible for the management and deployment of SFCs in the different life-cycle stages of SFCs. In this paper, we only present two applications, while others can be introduced to enhance the SFC flexibility and utilities. The SFC composition and path calculation modules are complementary. They collaborate for preparing a VNF forwarding graph to be deployed according to the requirements described in the SFC abstraction layer.

   *a) SFC instance composition:* The SFC composition module is responsible for translating the SFC blueprint to a concrete SFC by selecting the accurate SF instances for an SFC and mapping them to the corresponding resources. This module interacts with the path calculation module to optimize the VNF forwarding graph; both VNF resources consumption and links properties should be used.

   *b) Symmetry-aware path calculation :* This module contains all the intelligence in calculating the accurate path for an SFC. Several criteria can affect the path choice, including the SLA definition and the QoS required by the intent request. We propose in section III.2 an algorithm to calculate the reverse path for an SFC taking into consideration the symmetry criterion of SFs.

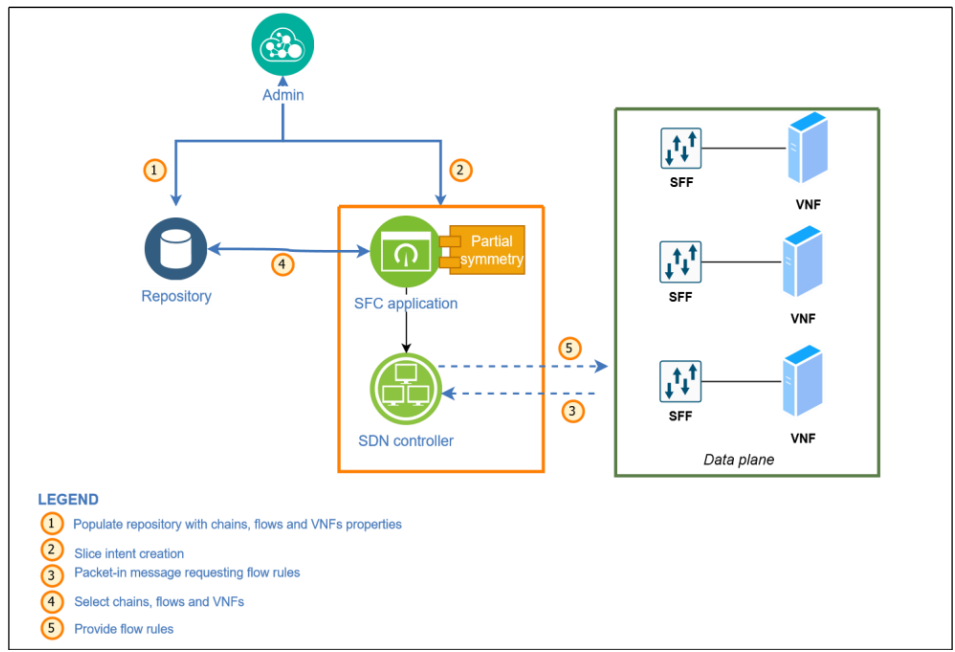

Fig. 4: System architecture for implementing the partial symmetry SFC.

## B. Reverse-path calculation algorithm

One of the main modules of the framework operations layer is the symmetry-aware path calculation module. As stated earlier, this module is responsible for dynamically calculating the accurate reverse and forward path for an SFC. In this section, we introduce the SFC reverse-path calculation algorithm. Its goal is to enhance the flexibility by dynamically programming and pushing flow rules responsible for traffic steering in the SFFs to ensure that the symmetry requirements of SFs are respected. Algorithm 1 describes the main steps of the reverse-path calculation. It first checks the traffic direction (forward/reverse), selects the appropriate SFC, and iterates for the consecutive SFs. Starting by the last SF (the term VNF is also used to refer to SF), an attribute characterizing the traffic symmetry is checked. A repository is used to store information about the SFCs, flows, paths and SFs. Once the symmetry attribute is true, a flow rule is added to the last SFF in the SFC to direct the reverse traffic to pass by that SF. Similarly, the symmetry of the other SFs is checked and the flow rules are added according to reverse-path information stored in the repository.

The traffic steering method used is based on Mac addresses. The MAC address of the next SF is inserted in the packets and modified along the path. If a flow should pass through SF1, the packets will have the destination MAC address modified to the SF1'MAC address. The last SFF in the path restores the original destination MAC address for the packets to be forwarded to the final destination. Each flow corresponds to a given SFC, while the reverse flow corresponds also to the same SFC, but passes through the SFs in the reverse order and counts only the SFs that are identified as "requiring symmetry".

**Algorithm 1: Symmetry-aware reverse path calculation**

Result: Reverse path

```
initialization;
if direction is reverse then
    if SFC_id is valid then
        if VNF is valid and available then
            select VNF_SFF();
            select in_port();
            if the VNF requires symmetry then
                add_flow(last SFF,
                    eth_dst=VNF_MAC);
            end
        end
        select prev_vnf();
        while true do
            if SFC_id is valid then
                if the VNF requires symmetry then
                    if SF is the last VNF then
                        add flow(last SFF, in_port,
                            eth_dst=VNF_MAC);
                    end
                else
                    add flow(SFF_SF,in_port,VNF_MAC);
                end
            end
            select VNF_SFF();
            select in_port();
        end
    end
end
select last VNF();
```

### IV. IMPLEMENTATION AND VALIDATION

In this section, we describe a partial symmetry proof of concept implementation and a preliminary performance analysis. The main goal is to estimate the gain in time by the partially symmetric chains compared to the fully symmetric chains.

### A. Implementation

We have developed an application to implement the Algorithm for calculating the reverse-path and the traffic steering method based on SFs symmetry requirements. The application is developed for Ryu SDN controller. It is an SFC prototype written in Python. Mainly, the application enforces OpenFlow forwarding rules for a defined flow. Such specific traffic is subject to an SFC and passes through a set of defined SFs. As a result of the application execution, a set of flow rules are installed in the SFFs to support the traffic steering operation. Fig. 4 presents the system architecture used to implement partial symmetry. First, an administrator populates the repository with information about flows, VNFs, and SFCs. Once the SFF does not find a rule to forward a particular flow, it sends a packet-in message to the SDN controller. Next, the SFC application interrogates the repository and calculates the forwarding and reverse-path, which is then communicated to the SFFs. Once the new forwarding rules are installed, the traffic can be steered according to the SFC symmetry required by the SFs.

*B. validation*

*1) Evaluation environment:* To evaluate the SFC application's performance, we use a data plane composed of Linux namespaces to emulate SFs and endpoints and OpenVswitch to implement the SFF and CL roles and OpenFlow to configure switches. The data plane components, along with the SDN controller and SFC application all run on an Ubuntu 18-64 bits equipped with 8G RAM, and Intel i7 processor using Python 3.6 interpreter. Fig.5, represents a simple topology to test the SFC symmetry concept. For the sake of simplicity, we choose to compare the partial symmetry concept with a full symmetry scenario, where the same instances of SFs have to be visited. In this scenario, the SFs do not treat traffic; they forward traffic to justify the SF traversal. Different interfaces are used in the SFs to precise the input and output traffic in the forward direction; the reverse traffic enters from the output direction and exits from the input direction. The chain used in the evaluation scenario imposes the following order:

$$SF2 =\Rightarrow SF1 =\Rightarrow SF3 \qquad (1)$$

In the partial symmetry scenario Fig.5a, SF1 and SF2 are identified as not requiring symmetry while SF3 requires symmetry. Thus, the reverse traffic, from the server to the client, should only enter the SF3. In contrast, Fig.5b represents the full symmetry scenario, where the reverse traffic should pass through all the SFs in the chain.

*2) Results discussion:* To test the performance of the SFC application and assess the added value of partial symmetry implementation, we compare a partial symmetry scenario (Fig.5a) with a full symmetry scenario (Fig.5b). The purpose of the experiments is to assess their performance according to the service delivery time including RTT, data transfer rate, and network throughput, respectively. We used iperf3, a traffic generator to generate the same amount of traffic in the two scenarios; Hence, we generate a flow of UDP datagrams, with a payload size equal to 1024 bytes. A total of 1 gigabyte is sent in the two scenarios, with a bandwidth equal to 1 gigabyte. The flow is sent from the server to the client to capture how the two scenarios react. The experiment is repeated 100 times for each scenario, the results reflect only the reverse traffic; they are captured at the client-side and are shown in Fig. 6a, Fig.6b and Fig. 6c with a confidence interval of 95%. To get RTT, we run the ping command, sending 100 ICMP requests, Fig.6a presents the RTT. We observe from the Fig.6a that in the partial symmetry scenario the RTT time is reduced, a result that can be attributed to the lower number of SFs that the reverse path passes through. Concerning data transfer, or the amount of data transferred per second, Fig.6b shows that the overall experiment time in the partial symmetry is less than the full symmetry case. This can be explained by the fact that the total amount of traffic is rapidly transferred in the partial symmetry scenario. We can see that the data transferred per second is higher in the partial symmetry scenario compared to the full symmetry. According to throughput variation (Fig. 6c), the throughput used is higher in the partial symmetry scenario compared to the full symmetry scenario, and this is because of reducing the number of SFs traversed in the reverse path. Thus the traffic does not have to pass by all the SFs, no unnecessary processing overhead is introduced, and hence better throughput is observed. To summarize, the partial symmetry SFC implementation reduces service delivery time and avoid extra load on the SFs.

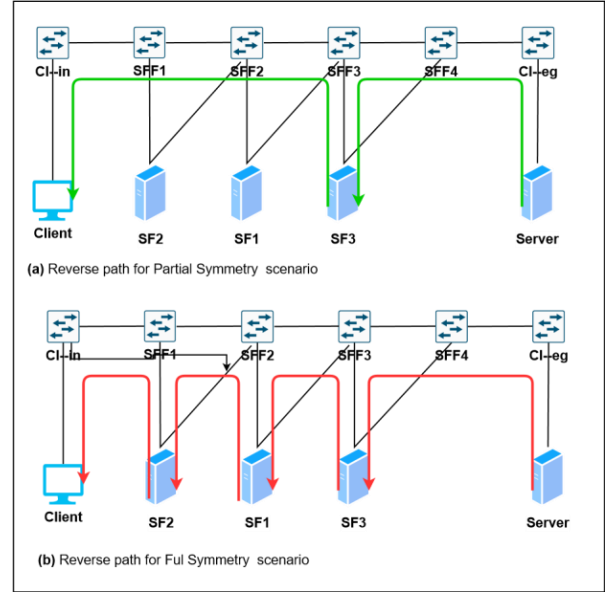

Fig. 5: Evaluation topology for SFC full and partial symmetry scenarios.

V. RELATED WORK

Thanks to the emergence of SDN and NFV technology, SFC has gained growing interest from both industry and academia. Though several studies have been published addressing various challenges of SFC [1], [5]–[7], very few works have considered traffic symmetry. Bifulco et al. in [8] have proposed CATENAE, which is an SFC system for mobile networks based on a ready to deploy traffic steering method. The proposed method considers SFs that modify the packet's header. The authors did not consider any SFC header-based or tunneling techniques that play a critical role in infrastructures that may not support new SFC headers. As for the symmetry issue, their traffic steering technique is based on NAT, used as the last SF. They switched the source and destination addresses and ports for the reverse direction to restore the flow header

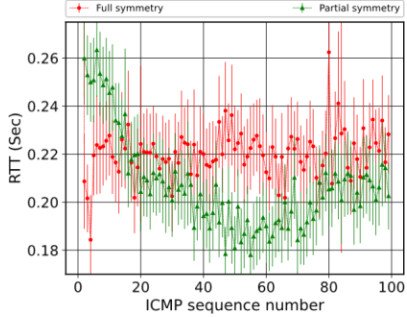

(a) Round trip time variation for full andpartial symmetry scenarios.

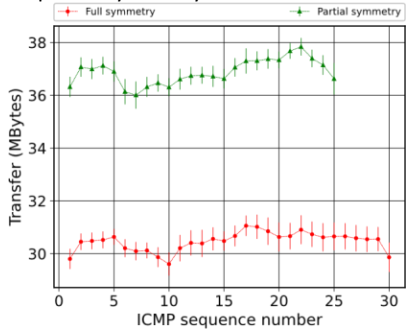

(b) Transfer variation in time for full andpartial symmetry scenarios.

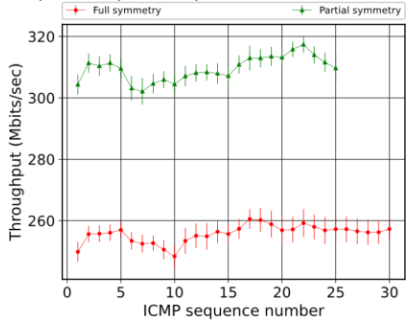

(c) Throughput variation in time for fulland partial symmetry scenarios.

Fig. 6: Testbed results for partial symmetry compared to full symmetry

in the forward direction. This work is among the very few works that considered the symmetry issue in SFC. However, this technique only permits two possible configurations: full symmetry or no symmetry in the chain. Hence there is no consideration of per-SF symmetry requirement.

Li et al. in [9] proposed an application-aware security SFC breaching approach, based on carrying the proactively analyzed application features in NSH metadata to be processed by security SFs. Their system mainly ensures that the dataplane can redirect traffic based on metadata, without the need for a control plane to participate. The authors defined a chain for each direction, where the SFs have two interfaces: an eastbound interface for the forward path and a westbound interface for the reverse path. Thus, the accurate SFC is selected according to the interface from where the traffic is coming along with the source and destination addresses. The traffic steering method used is based on NSH, this SFC protocol is being supported by many SDN controllers and Openflow switches to enable dynamic SFC. However, for the large scale use cases, the number of SFCs can be limited by consuming two SFC identifiers for each traffic flow. Also, they did not consider the individual SFs symmetry requirements. The SFC is either bidirectional or unidirectional(no symmetry). A patent also tackled the problem of partial symmetry that concerns the reverse-path [10]. Authors in [10] assumed that SFCs are unidirectional by default, but some SFs are stateful, such as the stateful IP services that require symmetry at the same SF instance. Thus, the reverse-path should only include the stateful SFs. The controller identifies such SFs and then generates information for the reverse path accordingly. The forward SFC-Id is bounded with a reverse SFC-Id. The main advantage of this method is the intelligence of detecting the SFs requiring symmetry by the controller and binding a forward chain with a reverse chain to include the stateful SFs. Yet, two SFC identifiers are used (per flow), which may not scale when the traffic steering method used is based on a small field to communicate the SFC id. Our proposed method binds the reverse traffic to the same SFC ID and calculates the reverse-path using a high-level application (on top of the control plane) to cope with such a problem.

In contrast to the previously mentioned solutions, our proposed framework considers the symmetry requirements of each SF in the SFC without keeping the state in the classifiers or implementing complex configurations. To this end, we use a repository interacting with the SFC application, and we proactively define the SFCs, the flows, and VNF properties (including the symmetry requirement attribute). The reverse symmetry algorithm programs the reverse path, and the accurate flow rules are then pushed to the SFFs and CLs. Therefore, our framework ensures partial symmetry cases while enhancing flexibility and automation.

## VI. Conclusion

To the best of the authors' knowledge, this is the first work to propose the deployment of the concept of partial symmetry in SFC for future 5G networks. This is expected to reduce the delivery time and the load on the SFs. The article has presented a framework for assisting the automation and administration of SFC, taking into consideration the partial symmetry issue. Furthermore, an SFC-aware path calculation algorithm is presented along with a prototype implementation and evaluation. This article shows that implementing partial symmetric SFCs in 5G networks can reduce delivery time, load on VFNs and the related cost and surely increase the network management flexibility. Moreover, such framework ensures cohabitability, and elasticity to meet the network requirements while reducing costs.

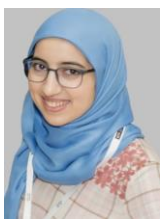
Hajar Hantouti received her M.S. degree in information systems security from the University of Ibn Tofail, Morocco, in 2014. She is currently a Ph.D. candidate at the University of Moulay Ismail, Morocco. Her research interests include Service Function Chaining, Software Defined Networking and Network Function Virtualization.

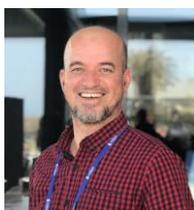
Nabil Benamar received his Master and PhD degrees from Moulay Ismail University of Meknes, Morocco, in 2001 and 2004 respectively. He is currently a professor of computer sciences at the School of Technology, Moulay Ismail University of Meknes, Morocco, and an Adjunct faculty of computer Sciences at Al Akhawayn University in Ifrane, Morocco. He is an IPv6 expert and consultant with many international organizations. His research interests include Service Function Chaining, vehicular networks, DTNs, ITS, IPv6 and IoT. He has authored or coauthored several journal articles in highly ranked journals and conferences. He has authored the RFC8691 and other IETF Internet documents.

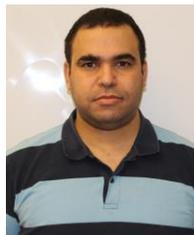
Miloud Bagaa received the engineer's, master's, and PhD degrees from the University of Science and Technology Houari Boumediene, Algiers, Algeria, in 2005, 2008, and 2014, respectively. From 2009 to 2015, he was a researcher with the Research Center on Scientific and Technical Information, Algiers. From 2015 to 2016, he was with the Norwegian University of Science and Technology, Trondheim, Norway. He is currently a senior researcher at Aalto University. His research interests include wireless sensor networks, Internet of Things, 5G Wireless communication, security, and networking modeling.

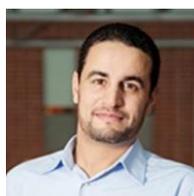
Tarik Taleb received the BE degree (Hons.) in information engineering and the MSc and PhD degrees in information sciences from GSIS, Tohoku University, Sendai, Japan, in 2001, 2003, and 2005, respectively. He is currently a professor at the School of Electrical Engineering, Aalto University, Espoo, Finland. He is a member of the IEEE Communications Society Standardization Program Development Board. In an attempt to bridge the gap between academia and industry, he founded the IEEE Workshop on Telecommunications Standards: From Research to Standards,a successful event that was recognized with the Best Workshop Award by the IEEE Communication Society (Com-SoC). Based on the success of this workshop, he has also founded and has been the Steering Committee Chair of the IEEE Conference on Standards for Communications and Networking. He is the general chair of the 2019 edition of the IEEE Wireless Communications and Networking conference held in Marrakech, Morocco. He is /was on the editorial board of the IEEE Transactions on Wireless Communications, IEEE Wireless Communications Magazine, IEEE Journal on Internet of Things, IEEE Transactions on Vehicular Technology, IEEE Communications Surveys and Tutorials, and a number of Wiley journals. He is also an IEEE Communications Society (ComSoc) distinguished lecturer.